\documentclass[3p,times]{elsarticle}

\usepackage{ecrc}

\volume{00}

\firstpage{1}

\journalname{Physics Letters B}

\runauth{Wojnar, Kalita \& Sarmah}

\jid{plb}

\jnltitlelogo{Physics Letters B}

\CopyrightLine{2023}{Published by Elsevier Ltd.}

\usepackage{latexsym}
\usepackage{amssymb, amsmath, bm, physics}
\usepackage{subcaption}
\usepackage{mathptmx}
\usepackage{flushend}
\usepackage[colorlinks,citecolor=blue,urlcolor=blue,linkcolor=blue]{hyperref}
\usepackage{aas_macros}
\usepackage{orcidlink}

\biboptions{sort&compress}

\begin{document}

\begin{frontmatter}



\dochead{}

\title{Effects of modified gravity on microscopic properties and cooling timescale of white dwarfs}


\author{Aneta Wojnar\orcidlink{0000-0002-1545-1483}}
\address{Departamento de F\'isica Te\'orica, Universidad Complutense de Madrid, E-28040, Madrid, Spain}
\ead{awojnar@ucm.es}

\author{Surajit Kalita\orcidlink{0000-0002-3818-6037}}
\address{High Energy Physics, Cosmology \& Astrophysics Theory (HEPCAT) Group, Department of Mathematics \& Applied Mathematics, \\ University of Cape Town, Cape Town 7700, South Africa}
\ead{surajit.kalita@uct.ac.za}

\author{Lupamudra Sarmah\orcidlink{0000-0003-1651-9563}}
\address{Indian Institute of Astrophysics, Bengaluru 560034, India}
\ead{lupamudra.sarmah@iiap.res.in}

\begin{abstract}
There are currently two open questions in white dwarf physics: why are massive dwarfs observed less often in astronomical surveys, and why have not any super-Chandrasekhar white dwarfs been found despite the discovery of more than a dozen peculiar, overly-luminous type Ia supernovae in about a couple of decades? According to different research, magnetic fields appear to somewhat resolve these issues, but stability remains a concern. For the first time, we investigate how modified gravity affects the specific heat of electrons and ions, the crystallization process, and the cooling mechanism in white dwarfs. We demonstrate it for the Ricci-based gravity. We show that massive white dwarfs fade faster and conclude that it could be a physical reason, apart from the presence of high magnetic fields, both for finding fewer massive white dwarfs and the lack of direct detection of super-Chandrasekhar white dwarfs.
\end{abstract}

\begin{keyword}
modified gravity \sep white dwarfs \sep Chandrasekhar limit \sep luminosity \sep specific heat

\end{keyword}

\end{frontmatter}


\newpage
Many proposals for the extensions of General Relativity (GR) have been proposed to shed light on the dark sector of the Universe, such as the mismatch between observations of the visible components in galaxies and their dynamical mass, as well as `too massive' compact objects~\cite{nojiri2017modified,saridakis2021cantata,abbott2020gw190521}. The modifications of GR seem unavoidable in the cosmological regimes while in small-scale systems such as compact objects and solar systems, they should be significantly suppressed. Nevertheless, the gravitational parameter space diagrams~\cite{baker2015linking} reveal noticeable untested regimes for the curvature values in which one can find galaxies, stellar objects, and also white dwarfs. These gaps, separating small-scale systems from cosmological ones, could potentially hide the onset of corrections to GR. 

As it turns out, the most popular alternate gravity theories, such as the scalar-tensor gravity and metric-affine theories, are not only suitable to explain at least some parts of the GR’s shortcomings, but they also modify their Newtonian limits~\cite{Olmo:2019flu}. Generally, the Poisson equation for such theories of gravity can be written as
    \begin{equation}\label{poisGen}
        \nabla^2\Phi = 4\pi G \rho +\alpha\,\text{\sc mgt},
    \end{equation}
where the modified gravity term $\text{\sc mgt}$ is a general function of density $\rho$ 
and their derivatives, whose exact form depends on the characteristics of a given theory of gravity while $\alpha$ is the model parameter. In a spherically symmetric spacetime with $r$ being the radial coordinate, $\text{\sc mgt}$ is often a function of $\rho(r)$ and its $r-$derivatives 
for the most popular gravity proposals \cite{toniato2020palatini,banados2010eddington,pani2011compact,koyama2015astrophysical}. In such a case, the hydrostatic balance and the mass estimate equations are respectively given by
\begin{align}\label{hydro0}
    \dv{\Phi}{r} = -\frac{1}{\rho}\dv{P}{r}\quad \text{and} \quad \dv{\mathcal{M}}{r} = 4\pi r^2 \rho,
\end{align}
where $\mathcal{M}(r)$ is the mass and $P(r)$ is the pressure, related to $\rho(r)$ by a barotropic equation of state.

A class of astrophysical objects which can be considered under the above formalism, are white dwarf stars. Various surveys, e.g., GAIA, SDSS, Kepler, etc. have explored white dwarfs, deriving surface temperatures and masses from measured fluxes and photometric distances\footnote{Catalogs are available at \url{http://www.astronomy.villanova.edu/WDCatalog/index.html} and \url{https://warwick.ac.uk/fac/sci/physics/research/astro/research/catalogues}.}. Notably, super-Chandrasekhar white dwarfs remain undetected, with most white dwarfs having masses below $1\,\mathrm{M_\odot}$. For an in-depth analysis of white dwarf mass distribution, refer to \cite{tremblay2016field,kepler2016white}, which highlights a scarcity of massive white dwarfs. The absence of super-Chandrasekhar white dwarfs in these surveys may be attributed to their potential high magnetization, leading to reduced luminosity \cite{ferrario2015magnetic,bhattacharya2018luminosity}. However, exceeding the Chandrasekhar limit requires an exceptionally strong magnetic field, possibly rendering the white dwarf unstable \cite{braithwaite2009axisymmetric}. Thus, modified gravity emerges as a strong plausible explanation.

While super-Chandrasekhar white dwarfs are not directly observed, their existences are indirectly predicted based on peculiar over-luminous type Ia supernovae. This study aims to demonstrate that modified gravity accelerates the cooling process for higher-mass white dwarfs, resulting in a diminished population as mentioned above~\cite{tremblay2016field,kepler2016white,ferrario2015magnetic,bhattacharya2018luminosity,braithwaite2009axisymmetric}.

In the further part of the paper, we focus only on the case where $\text{\sc mgt}$ is a function of $\rho$ and its derivatives only \cite{olmo2021parameterized}, which is a correction term to the Poisson equation resulting from the Ricci-based theory of gravity \cite{alfonso2017trivial,afonso2018mapping}. Theories such as GR ($\alpha=0$) but also Eddington-inspired Born-Infeld ($\alpha\rightarrow \kappa^2\alpha/4 $) \cite{,banados2010eddington,pani2011compact} and Palatini $f(R)$ ($\alpha\rightarrow 2\kappa^2\alpha $) \cite{toniato2020palatini} are the most popular examples. Note that for the last two cases, we have $\text{\sc mgt}=\nabla^2\rho$. Those GR extensions, with Lagrangian functionals built of the symmetric part of the Ricci tensor contracted with metric, are considered \`a la Palatini: the metric and affine connection are independent variables. This feature does not only provide the second-order field equations (thus, no extra degree of freedom apart from the GR polarization of the gravitational field), but also that in the vacuum and radiation-dominated eras, the field equations reduce to the GR ones with a cosmological constant. However, the non-vacuum case differs with respect to Einstein's gravity, therefore providing different scenarios in early and late-time cosmology, as well as in astrophysics.

The most stringent constraint on the parameter $\alpha$ is presently provided in~\cite{Kozak:2023axy,Kozak:2023ruu,Wojnar:2023bvv}. Specifically, the current precision confines $\alpha$ within the range of $-2 \times 10^9$ to $10^9 \rm\,m^2$ with a 2$\sigma$ accuracy, while ensuring microscopic stability provides  $\alpha > -7.52\times 10^7 \text{ m}^2$. The latest cosmological data, available in~\cite{Gomes:2023xzk}, yields bounds approximately 40 orders of magnitude larger.

For the considered Ricci-based gravities, the hydrostatic equilibrium equation \eqref{hydro0} takes the form~\cite{2023PhRvD.107d4072K}\footnote{The hydrostatic equilibrium equation similarity in both gravity proposals is not accidental. In the 1st order approximation, EiBI gravity becomes Palatini gravity with a quadratic term. However, only the quadratic term affects the non-relativistic equations, as higher-order curvature scalar terms enter the equations at the sixth order \cite{toniato2020palatini}.}
\begin{equation}\label{hydro}
    \dv{P}{r}= -\frac{G\mathcal{M}\rho}{r^2} + 8\pi G \alpha \rho \dv{\rho}{r}.
\end{equation}
Because our target is to study internal properties and processes happening in white dwarfs, whose cores are predominantly comprised of degenerate electrons, they effectively follow the Chandrasekhar equation of state ~\cite{1935MNRAS..95..207C}
\begin{equation}
\begin{aligned}\label{Chandra}
    P &= \frac{\pi m_\text{e}^4 c^5}{3 h^3}\left[x_\text{F}\left(2x_\text{F}^2-3\right)\sqrt{x_\text{F}^2+1}+3\sinh^{-1}x_\text{F}\right],\\
    \rho &= \frac{8\pi \mu_\text{e} m_\text{p}(m_\text{e}c)^3}{3h^3}x_\text{F}^3,
\end{aligned}
\end{equation}
where $x_\text{F} = p_\text{F}/m_\text{e}c$ with $p_\text{F}$ being the Fermi momentum while other constants have their usual meaning.

The thermal energy of a spherical-symmetric object is given by
\begin{equation}\label{tenergy}
    U=\bar{c}_v\frac{\mathcal{M}}{A m_\mathrm{p}} T,
\end{equation}
where $\mathcal{M}/(Am_\mathrm{p})$ is the number of ions with $A$ being the mean atomic weight and $T$ the temperature of the isothermal core while the mean specific heat, taken for the whole stellar configuration, is as follows:
\begin{equation}\label{mean}
    \bar{c}_v=\frac{1}{\mathcal{M}}\int_0^\mathcal{M}(c_v^\text{el}+c_v^\text{ion})\dd{m}.
\end{equation}
The specific heat of the electrons per ion $c_v^\text{el}$ is given by~\cite{koester1972outer}
\begin{equation}
    c_v^\text{el}=\frac{3}{2}\frac{k_\mathrm{B}\pi^2}{3}Z \frac{k_\mathrm{B}T}{\epsilon_\mathrm{F}},
\end{equation}
where $Z$ is the charge and $\epsilon_\mathrm{F}=\sqrt{ p_\mathrm{F}^2c^2 + m_\mathrm{e}^2c^4}$ is the Fermi energy. The specific heat of ions $ c_v^\text{ion}$ takes the form 
\begin{equation}
    c_v^\text{ion}=9k_\mathrm{B} \left(\frac{T}{\Theta_\mathrm{D}}\right)^3 \int_0^{\Theta_\mathrm{D}/T} \frac{x^4 e^x}{\left(e^x-1\right)^2}\dd{x},
\end{equation}
where the Debye temperature is given by
\begin{equation}\label{debye}
    \Theta_\mathrm{D}=0.174\times10^4 \frac{2Z}{A}\sqrt{\rho}.
\end{equation}
Note that $\Theta_\mathrm{D}$ and both specific heats depend on the density profile $\rho(r)$ which is a solution of the hydrostatic equilibrium equations \eqref{hydro0}. 
From Eq.~\eqref{mean}, it is evident that (specific) heat capacity is not only a property of the material depending on temperature and state of matter, but it also depends on the model of gravity. Moreover, the mean specific heat in Eq.~\eqref{mean} contains phonons' contribution resulting from crystallization processes. Because the elastic and thermodynamic properties of crystals are expressed via $\Theta_\mathrm{D}$ as shown in Eq.~\eqref{debye}, which also relies on $\Theta_\mathrm{D} \propto \sqrt{\rho(r)}$, the characteristics of the crystal vary in different gravitational proposals.



\begin{figure}[htpb]
        \centering
	\includegraphics[scale=0.5]{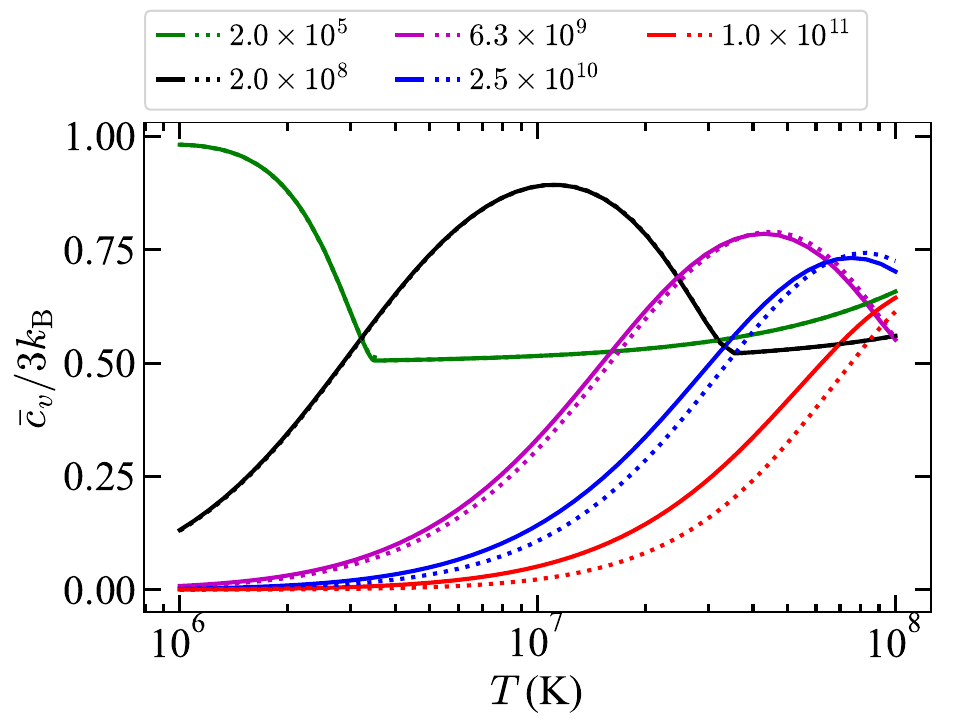}
	\caption{$\bar{c}_v$ as a function of $T$ for different carbon white dwarfs with their central densities are shown in the label in g\,cm$^{-3}$ units. Solid lines represent conventional white dwarfs under Newtonian gravity while dotted lines represent white dwarfs under modified gravity with $\alpha=-2\times10^{13}\rm\,cm^2$.}
	\label{Fig: CV_T}
\end{figure}

Fig.~\ref{Fig: CV_T} shows the variation of $\bar{c}_v$ as a function of $T$ for carbon white dwarfs. As expected, the effect related to gravity is more prominent at high densities. Since in modified gravity, the mean specific heat reaches lower values than in the Newtonian case for the same temperature, as we will see later, compact objects cool down faster with respect to Newton's gravity. Let us comment that in the case of a different composition, for instant O-Ne or O-Ne-Mg white dwarfs, they do not change the mass--radius relation significantly (see e.g. \cite{camenzind2007compact}). In this work, since our target is to understand the cooling age, we stick to one specific type of white dwarf as the overall final result does not alter significantly with the elements present.

The luminosity, which is determined by the rate of decrease in thermal energy of ions and electrons over time $t$, using Eq.~\eqref{tenergy}, can be expressed in the following form:
\begin{equation}\label{LU}
    L_\text{thermal}=-\dv{U}{t}=-\frac{\mathcal{M}}{A m_\mathrm{p}}\bar{c}_v\dv{T}{t}.
\end{equation}
Apart from it, we need to consider the latent heat released during the crystallization process. Assuming it as $q k_\mathrm{B} T$, the additional contribution to the total luminosity takes the form~\cite{van1968crystallization}
\begin{equation}
    L_\text{latent}=qk_\mathrm{B}T \dv{(m_s/A m_\mathrm{p})}{t},
\end{equation}
where $m_s$ is the amount of mass that is already crystallized. Therefore, the total luminosity $L =L_\text{thermal}+L_\text{latent}$ is given by
\begin{equation}
  L  = \frac{3k_\mathrm{B}\mathcal{M}}{A m_\mathrm{p}} \left(-\frac{\bar{c}_v}{3k_B} + \rho_s q \frac{1}{\mathcal{M}}\dv{m}{r}\dv{r}{\rho} \right)\dv{T}{t},
\end{equation}
where $\rho_s(T)$ is the density of the crystallized mass at a temperature $T$. It is derived from the ratio of Coulomb to thermal energy $\Gamma$ when it reaches a critical value for which the crystallization process starts~\cite{koester1972outer}
\begin{equation}
    \Gamma:=2.28\times10^5 \frac{Z^2}{A^{1/3}} \frac{\rho_s^{1/3}}{T} = \Gamma_\text{critical}.
\end{equation}

\begin{figure}[t]
	\centering
	\includegraphics[scale=0.5]{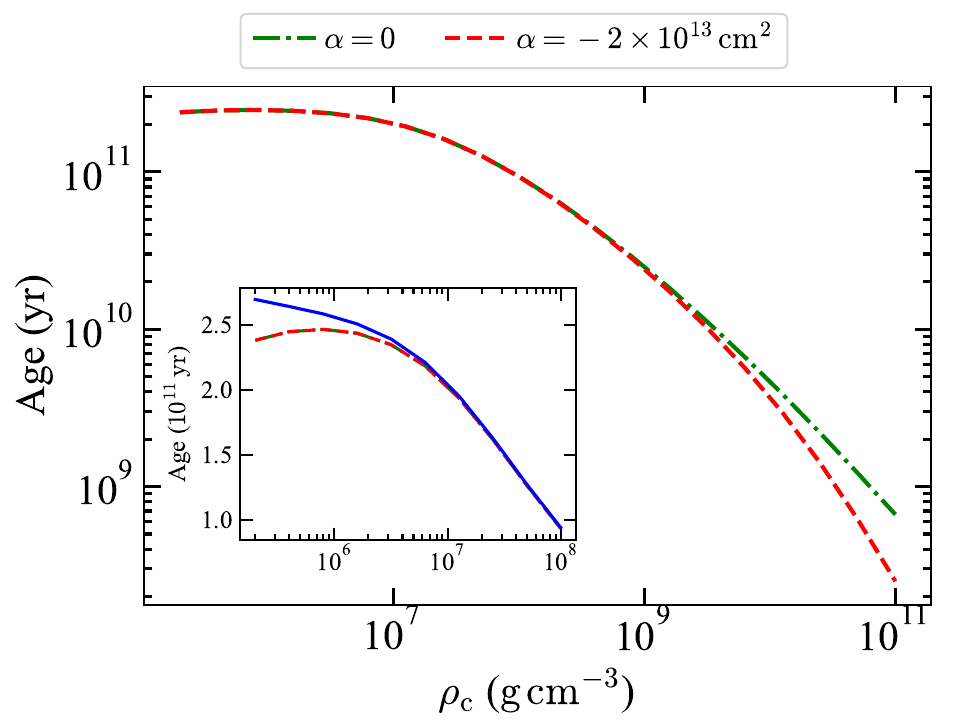}
	\caption{Age of carbon white dwarfs as a function of their central densities, obtained by solving Eq.~\eqref{LU} as they cool down from $10^8$\,K to $10^6$\,K. The blue curve in the inside plot shows the effect of crystallization on the low-mass white dwarfs.}
	\label{Fig: age1}
\end{figure}

\begin{figure}[t]
	\centering
	\includegraphics[scale=0.5]{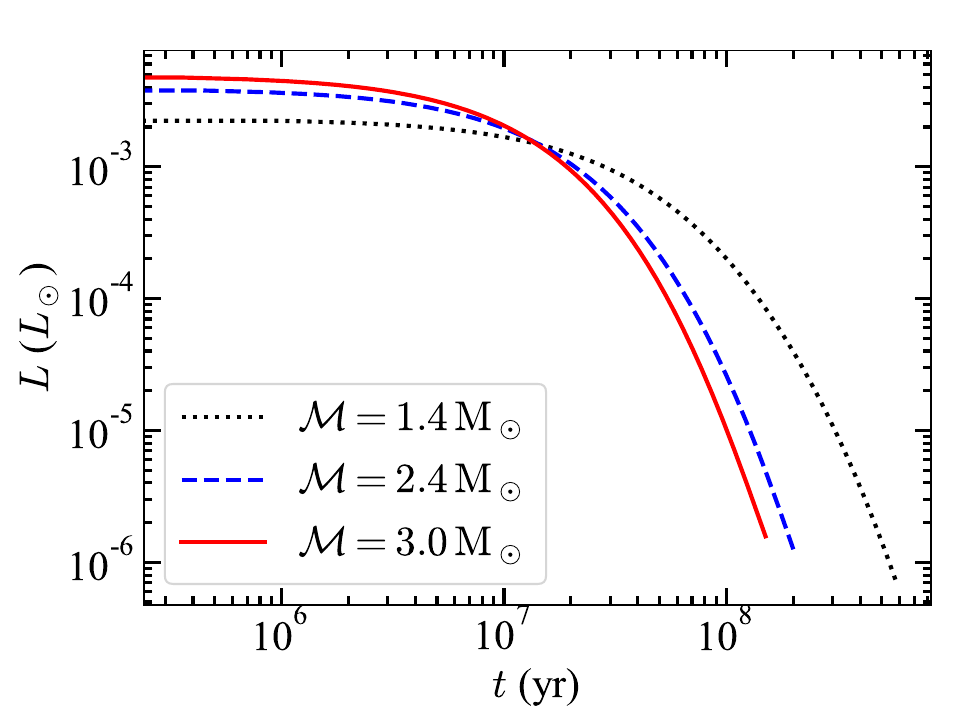}
	\caption{Change in $L$ with time for carbon white dwarfs assuming the surface temperature to be $10^7\rm\,K$ at $t=0$.}
	\label{Fig: Luminosity}
\end{figure}

Assuming the initial (i.e. at $t=0$) temperature to be $10^8$\,K, Fig.~\ref{Fig: age1} shows a carbon white dwarf's age (we understand the `age of a white dwarf' as cooling time from this temperature to the present values, say $10^6$\,K). Simultaneously, Fig.~\ref{Fig: Luminosity} shows the fading of this white dwarf with time and thereby the effect of modified gravity. It is evident that massive white dwarfs fade faster. A detailed discussion on this phenomenon in the context of modified gravity can be found in~\cite{2023PhRvD.107d4072K,2022Univ....8..647K}. This could be another reason for detecting less number of massive white dwarfs in astronomical surveys like GAIA, SDSS, etc.~\cite{2018MNRAS.480.4505J,2020ApJ...898...84K}, and also for the non-detection of super-Chandrasekhar white dwarfs directly so far. This idea is very different from those in the existing literature based on magnetic fields~\cite{2015SSRv..191..111F,2018MNRAS.477.2705B}, where it was shown that the surface luminosity of white dwarfs decreases with the increase in the magnetic field strength. Note that the presence of latent heat in the form of extra energy resulting from crystallization makes a longer cooling process as this energy must also be dissipated from the surface of the star, which is independent of the gravity theory as evident in Fig.~\ref{Fig: age1}. On the other hand, the cooling process is speeded up by modified gravity as illustrated in Fig.~\ref{Fig: Luminosity}. This is desirable for explaining the existence of white dwarfs that appear to be older than the universe, especially because of the fact that this effect is particularly noticeable in low-mass white dwarfs~\cite{masuda2019self}. Despite the inadequacy of established scenarios to account for this unusual phenomenon~\cite{laughlin1997end}, Ricci-based as well as other gravitational proposals may also offer viable solutions.

This is the first time gravity's influence on specific heat, Debye temperature, and crystallization process has been reported in the literature. This property can have a significant meaning in solid-state physics and Earth science. Recent experiments recreated extreme conditions of the Earth's core, allowing the study of iron's behavior under high pressure and temperature~\cite{merkel2021femtosecond}. While modified effects can be ignored in weak fields like Earth's laboratories, they are important in Earth and stellar interiors~\cite{Kozak:2021fjy,Benito:2021ywe,Wojnar:2022dvo}. Our findings can be used to understand gravitational interaction in dense environments~\cite{Kozak:2023axy} and test existing proposals of gravitational theory against white dwarf data \cite{jain2016white,Kalita:2023hcl}.

\section*{Acknowledgements}
We would like to thank the anonymous reviewer for their useful suggestions to improve the quality of the manuscript.
A.W. acknowledges financial support from MICINN (Spain) {\it Ayuda Juan de la Cierva - incorporac\'ion} 2020 No. IJC2020-044751-I. S.K. would like to acknowledge support from the South African Research Chairs Initiative of the Department of Science and Technology and the National Research Foundation.

\bibliographystyle{elsarticle-num}
\bibliography{bibliography}

\end{document}